\documentclass{emulateapj}

\newcommand{\Msun}{ M_{\odot}}
\newcommand{\Mdot}{\dot{M}}
\newcommand{\teff}{T_{\rm eff}}

\begin{document}
\title{The Cool Accretion Disk in ESO 243-49 HLX-1: Further Evidence of an Intermediate Mass Black Hole}
\author{Shane W. Davis\altaffilmark{1}, Ramesh Narayan\altaffilmark{2},
Yucong Zhu\altaffilmark{2},
Didier Barret\altaffilmark{3,4}, Sean A. Farrell\altaffilmark{5,6}, 
Olivier Godet\altaffilmark{3,4}, Mathieu Servillat\altaffilmark{2}, 
and Natalie A. Webb\altaffilmark{3,4}}
\altaffiltext{1}{Canadian Institute for Theoretical Astrophysics. Toronto, ON M5S3H4, Canada}
\altaffiltext{2}{Harvard-Smithsonian Center for Astrophysics, 60
  Garden Street, Cambridge, MA 02138}
\altaffiltext{3}{Universit\'e de Toulouse; Universit\'e Paul Sabatier - Observatoire
Midi-Pyr\'en\'ees, Institut de Recherche en Astrophysique et 
Plan\'etologie (IRAP), Toulouse, France}
\altaffiltext{4}{Centre National de la Recherche Scientifique; IRAP; 9 Avenue
du Colonel Roche, BP 44346, F-31028 Toulouse Cedex 4, France}
\altaffiltext{5}{Department of Physics and Astronomy, University of Leicester, University Road, Leicester, LE1 7RH, UK}
\altaffiltext{6}{Sydney Institute for Astronomy, School of Physics A29, The University of
Sydney, NSW 2006, Australia}

\begin{abstract}

With an inferred bolometric luminosity exceeding $10^{42}$~$\rm erg \;
s^{-1}$, HLX-1 in ESO 243-49 is the most luminous of ultraluminous X-ray
sources and provides one of the strongest cases for the existence of
intermediate mass black holes.  We obtain good fits to disk-dominated
observations of the source with BHSPEC, a fully relativistic black
hole accretion disk spectral model.  Due to degeneracies in the model
arising from the lack of independent constraints on inclination and
black hole spin, there is a factor of 100 uncertainty in the best-fit
black hole mass $M$. Nevertheless, spectral fitting of {\it
  XMM-Newton} observations provides robust lower and upper limits with
$3000 \Msun \lesssim M \lesssim 3\times 10^5 \Msun$, at 90\%
confidence, placing HLX-1 firmly in the intermediate-mass regime. The
lower bound on $M$ is entirely determined by matching the shape and
peak energy of the thermal component in the spectrum.  This bound is
consistent with (but independent of) arguments based solely on the
Eddington limit.  Joint spectral modelling of the {\it XMM-Newton}
data with more luminous {\it Swift} and {\it Chandra} observations
increases the lower bound to $6000 \Msun$, but this tighter constraint
is not independent of the Eddington limit.  The upper bound on $M$ is
sensitive to the maximum allowed inclination $i$, and is reduced to $M
\lesssim 10^5 \Msun$ if we limit $i \lesssim 75^\circ$.

\end{abstract}


\section{Introduction}

Ultraluminous X-ray sources (ULXs) are extragalactic X-ray sources
whose luminosities match or exceed the Eddington luminosity for
accretion onto a 10 $\Msun$ black hole (BH). It is thought that the
most luminous objects in this class may harbor BHs with masses in the
$100$ to $10^5 \Msun$ range (intermediate mass black holes or IMBHs).
BHs in this mass range are of particular interest because our current
understanding of stellar evolution suggests they could not be formed
by the collapse of a single massive star in the current epoch of star
formation \citep[e.g.,][and references therein]{bel10}.  The strongest
support for the IMBH interpretation is founded on theoretical
arguments that BH accretion flows cannot radiate significantly above
the Eddington luminosity.  If this argument holds, the most luminous
ULX sources, which exceed $10^{40}$~$\rm erg \; s^{-1}$, must host
IMBHs.  Alternatively, if real accretion flows are capable of
radiating significantly above the Eddington limit or their emission is
strongly beamed \citep{kin01}, these sources might contain BHs of only
a few tens of solar masses.  BHs in this lower mass range have been
identified through dynamical studies of X-ray binaries in the Milky
Way and other Local Group galaxies \citep{rm06}.  These sources can
easily be explained by the theory of stellar evolution and collapse.

Cool thermal components found in the spectra of some ULX sources
provide additional support for the IMBH interpretation.  Due to the
scaling of the characteristic gravitational radius and luminosity with
mass, accretion disks radiating at a fixed fraction of the Eddington
luminosity will tend to have lower maximum effective temperatures for
larger masses if the inner radius of the accretion flow scales with
the gravitational radius of the BH \citep{ss73,nt73}.  In many
luminous ULXs, fits with multicolor disk blackbody models favor
relatively cool disks \citep[e.g.,][]{mil03,mfm04a,mfm04b}.  However,
the disk component in many of these spectral fits only accounts for a
small or modest fraction of the bolometric power \citep[see
  e.g.][]{sd06,sor09}, which is instead dominated by a power law
component, generally thought to be inverse Compton scattering of disk
photons by hot electrons.  If the majority of the emission originates
in the hard component, it is no longer clear that the thermal
component is associated with emission from near the inner edge of the
disk (as opposed to larger radii), and the argument that a larger
emitting area follows from a larger gravitational radius (i.e. a large
mass) is weakened.

Making a convincing argument for an IMBH on the basis of spectral fits
requires observations in which a thermal component dominates the
bolometric emission.  However, such observations appear to be rare for
ULXs with luminosities $\gtrsim 10^{40}$~$\rm erg \; s^{-1}$
\citep{sd06,ber08,sk08,fk09,sor09}.  Most luminous ULXs are observed
in a power-law dominated state resembling the hard state or steep
powerlaw state of BH X-ray binaries \citep{mr06}, although there are
also suggestions that this may be a new mode of super-Eddington
accretion \citep[e.g.,][]{sd06,grd09}. Exceptions may include recent
observations of M82 X-1 \citep{fk10} and HLX-1 in ESO 243-49,
hereafter referred to simply as HLX-1 \citep{far09}, in which the
spectra of these sources appear to be dominated by a thermal
component.

In this work we focus on HLX-1, an off-nuclear X-ray source in the
galaxy ESO~243-49, which reaches luminosities in excess of
$10^{42}$~$\rm erg \; s^{-1}$ \citep{far09,god09,web10}.  The source
has been observed a number of times with {\it XMM-Newton}, {\it
  Swift}, and {\it Chandra}, displaying long term spectral variability
that is consistent with state transitions in Galactic X-ray binaries
(\citealt{god09}; Servillat et al., in preparation).  This large
luminosity is contingent on its placement in ESO~243-49.
\citet{sor10} have argued that the X-ray spectrum could plausibly be
explained as a foreground neutron star in the Milky Way.  However,
based on recent spectroscopic observations with the Very Large
Telescope, \citet{wie10} identify H$\alpha$ emission coincident with
HLX-1 at a very similar redshift to that of the host galaxy, placing
the source firmly in ESO~243-49.

For fitting the spectrum of HLX-1, we use BHSPEC \citep{dav05,dh06}, a
relativistic accretion disk spectral model, which has been implemented
as a table model in XSPEC \citep{arn96}.  The major advantage of this
model is that the spectrum of the emission at the disk surface is not
assumed to be blackbody, but is instead computed directly using TLUSTY
\citep{hl95}, a stellar atmospheres code which has been modified to
model the vertical structure of accretion disks \citep{hub90,hh98}.
The BH mass (and spin) are parameters of the model and are directly
constrained by spectral fitting.

BHSPEC, alone or in concert with KERRBB \citep{li05}, has been used to
fit numerous Galactic BH X-ray binaries.  For observations in which
the thermal component dominates, it provides a good fit and reproduces
the spectral evolution as luminosity varies \citep[e.g.,][and
  references therein]{sha06,ddb06,ste10}.  However, the mass range
covered by the BHSPEC models used in those works was below the IMBH
range.  BHSPEC models in the IMBH regime were generated by
\citet{hkh05}.  With these models, \citet{hk08} obtained good fits and
found best-fit BH masses in the 23-73 $\Msun$ range for five of the six
ULX sources in their sample, all of which were at least an
order-of-magnitude lower in luminosity than HLX-1. We have extended our
own version of the BHSPEC model into the IMBH regime and have used
these models for the analysis presented here, but the methods are
identical to those employed by \citet{hk08}.

This work is organized as follows:  We briefly describe the BHSPEC
model and discuss our data selection in Section~\ref{meth}.  We
summarize the main results of our spectral analysis in
Section~\ref{res}, and provide a more thorough discussion of our
constraints in Section~\ref{disc}.  We summarize and conclude in
Section~\ref{conc}.

\section{Methods}
\label{meth}

\subsection{Spectral Models}

In this work, we focus exclusively on the BHSPEC model.  We are
motivated by the observational evidence of a thermally dominant soft
X-ray component in the spectra of HLX-1 considered here and the
similarity of these spectra to those observed in Galactic BH X-ray
binaries.  Hence, we assume from the outset that the emission is from
a radiatively efficient, thin accretion disk, and derive constraints
on the parameters of interest, most importantly the BH mass, $M$.  We
refer readers interested in comparisons with other models to Godet et
al. (in preparation), which fits a wider array of models and attempts
to differentiate between various interpretations.

In addition to $\log M$, the BHSPEC model has 4 fit parameters:
a normalization, the BH spin $a_*$, the cosine of the inclination $i$,
and the log of the Eddington ratio $\ell = L/L_{\rm Edd}$, where
$L_{\rm Edd}=1.3 \times 10^{38} (M/\Msun) \rm \; erg \; s^{-1}$ is the
Eddington luminosity.  The normalization can be fixed using the known
distance to the source of $D \sim 95$~Mpc, leaving 4 parameters to fit.
We have computed models for $\log \ell = -1.5$ to 0, $\log
M/\Msun=3.25$ to 5.5, $\cos i =0$ to 1, and $a_* =-1$ to 0.99.  We assume
a \citet{ss73} stress parameter $\alpha=0.01$, but discuss the
implications of this choice in Section~\ref{disc}.

Although the spectra under consideration are thermally dominated,
there is a tail of emission at high energies which is not accounted
for by BHSPEC.  We model this emission with SIMPL \citep{ste09b},
which adds two free parameters: the power law index $\Gamma$ and the
fraction of scattered photons $f_{\rm sc}$ relative to the BHSPEC
model.  Photoelectric absorption by neutral gas along the line of
sight is accounted for by the PHABS model in XSPEC.  We leave the neutral
hydrogen column $N_{\rm H}$ free, adding a single additional
parameter.

\begin{figure}
\includegraphics[width=0.5\textwidth]{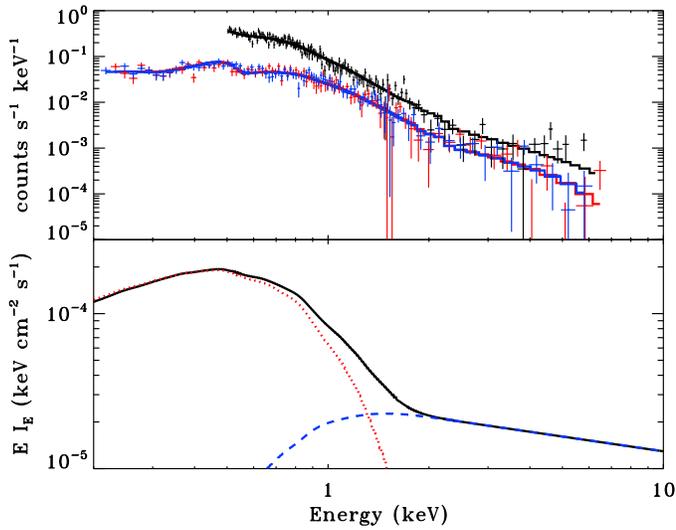}

\caption{XMM2 spectrum (top) and best-fit unabsorbed model (bottom)
  versus energy for $\cos i=1$ in Table~\ref{t:xmm}.  The top panel
  shows the observed count rate and best-fit model (with absorption)
  for the EPIC pn (black), MOS1 (red) and MOS2 (blue) data.  The bottom
  panel shows the best-fit total model (solid, black), BHSPEC
  component (dotted, red), and SIMPL component (dashed, blue) with
  absorption removed.
\label{f:spec}}
\end{figure}

\begin{deluxetable*}{cccccccc}
\tablecolumns{8}
\tablecaption{Best-Fit{\it XMM-Newton} Parameters\label{t:xmm}}
\tablehead{
\colhead{$\cos i$} &
\colhead{$a_*$} &
\colhead{$\log M$\tablenotemark{a}} &
\colhead{$\log \ell$} &                                                             
\colhead{$N_H$\tablenotemark{b}} &
\colhead{$\Gamma$} &
\colhead{$f_{\rm sc}$} &
\colhead{$\chi^2$/d.o.f.}
}
\startdata
$0$ &
$0.23^{+0.42}_{-0.04}$ &
$4.79^{+0.10}_{-0.05}$ &
$0.00^{+0}_{-0.22}$ &
$1.5^{+1.2}_{-1.0}$ &
$2.37^{+0.39}_{-0.28}$ &
$0.108^{+0.044}_{-0.027}$ &
$332/328$ \\
$0.25$ &
$0.99^{+0}_{-0.07}$ &
$5.07^{+0.07}_{-0.20}$ &
$-1.19^{+0.27}_{-0.01}$ &
$2.6^{+1.0}_{-1.2}$ &
$2.72^{+0.40}_{-0.36}$ &
$0.142^{+0.072}_{-0.043}$ &
$339/328$ \\
$0.5$ &
$0.99^{+0}_{-0.04}$ &
$4.78^{+0.08}_{-0.19}$ &
$-0.97^{+0.09}_{-0.04}$ &
$2.9^{+1.2}_{-1.2}$ &
$2.54^{+0.39}_{-0.46}$ &
$0.108^{+0.053}_{-0.041}$ &
$340/328$ \\
$0.75$ &
$-0.93^{+1.92}_{-0.07}$ &
$3.72^{+0.96}_{-0.04}$ &
$-0.16^{+0.02}_{-0.56}$ &
$2.5^{+1.2}_{-1.0}$ &
$2.28^{+0.45}_{-0.35}$ &
$0.080^{+0.030}_{-0.024}$ &
$338/328$ \\
$1$ &
$0.51^{+0.48}_{-1.51}$ &
$3.83^{+0.31}_{-0.32}$ &
$-0.37^{+0.22}_{-0.09}$ &
$2.5^{+1.1}_{-1.0}$ &
$2.31^{+0.39}_{-0.39}$ &
$0.085^{+0.044}_{-0.035}$ &
$336/328$ \\
$0.94^{+0.06}_{-0.06}$ &
$-1$ &
$3.61^{+0.18}_{-0.06}$ &
$-0.16^{+0.03}_{-0.02}$ &
$1.9^{+1.3}_{-1.1}$ &
$2.37^{+0.41}_{-0.37}$ &
$0.091^{+0.050}_{-0.029}$ &
$336/328$ \\
$0.03^{+0.04}_{-0.00}$ &
$0$ &
$4.70^{+0.08}_{-0.11}$ &
$0.00^{+0}_{-0.07}$ &
$2.1^{+1.2}_{-0.9}$ &
$2.38^{+0.43}_{-0.33}$ &
$0.099^{+0.043}_{-0.022}$ &
$334/328$ \\
$0.00^{+0.03}_{-0}$ &
$0.7$ &
$4.89^{+0.08}_{-0.07}$ &
$-0.26^{+0.03}_{-0.08}$ &
$1.9^{+1.3}_{-1.1}$ &
$2.34^{+0.36}_{-0.34}$ &
$0.099^{+0.037}_{-0.026}$ &
$335/328$ \\
$1.00^{+0}_{-0.16}$ &
$0.95$ &
$4.07^{+0.21}_{-0.07}$ &
$-0.43^{+0.03}_{-0.15}$ &
$2.8^{+1.1}_{-1.0}$ &
$2.24^{+0.40}_{-0.39}$ &
$0.073^{+0.038}_{-0.025}$ &
$337/328$ \\
\enddata
\tablenotetext{a}{$M$ is in units of $\Msun$.}
\tablenotetext{b}{$N_{\rm H}$ is in units of $10^{20} \; cm^{-2}$.}
\tablecomments{
Parameters tabulated without errors were fixed during the fit.  Errors
are computed using $\Delta \chi^2=2.706$ (90\% confidence for one
parameter).  Joint confidence contours are shown in Figures~\ref{f:spin}
 and \ref{f:inc}.}
\end{deluxetable*}

\subsection{Data Selection}
\label{data}

We begin with a collection of prospective disk dominated observations
of HLX-1.  This includes two observations with {\it XMM-Newton} (2004
November 23: XMM1; 2008 November 28: XMM2), two observations with {\it
  Swift} (2008 November 24: S1; 2010 August 30: S2), and one
observation with {\it Chandra} (2010 September 6).

The {\it XMM-Newton} observations were performed with the three EPIC
cameras in imaging mode with the thin filter. The two MOS cameras were
in full-frame mode for both observations.  The pn was operated in full
frame mode for the first observation (Obs-Id: 0204540201) and small
window mode for the second (Obs-Id: 0560180901).  The data were
reduced using the {\it XMM-Newton} Science Analysis System (SAS) v8.0
and event files were processed using the epproc and emproc tools. The
event lists were filtered for event patterns in order to maximize the
signal-to-noise ratio against non X-ray events, with only calibrated
patterns (i.e. single to double events for the pn and single to
quadruple events for the MOS) selected. Events within a circular
region of radius 60'' around the position of HLX-1 were extracted from
the MOS data in both observations and from the pn data in the first
observation. Background events were extracted for the same data from
an annulus around the source position with an inner radius of 60'' and
an outer radius of 84.85''. As the pn was in the small window mode
during the second observation, a smaller circular extraction region
with a radius of 40'' was used for extracting source
events. Background events were in turn extracted from a circular
region with radius 40'' from a region the same distance from the
center of the chip as HLX-1, which appeared to have a similar
background level in the image. Source and background spectra were
extracted in this way for each camera, with response and ancillary
response files generated in turn using the SAS tools rmfgen and
arfgen.

Although the MOS and pn spectra are consistent down to $\sim0.5$~keV,
below this boundary the pn spectrum deviates significantly, showing a
sharp, absorption-like feature that is absent in the MOS spectra.
Since this indicates a problem with the low energy response of the pn
camera in this observation, we follow \cite{far09} and ignore channels
with energies below 0.5 keV when fitting the pn spectrum.
For the MOS spectra we fit channels with energies greater than
0.2~keV.

The {\it Swift} XRT data were processed using the tool XRTPIPELINE
v0.12.3.4, as discussed in \citet{god09}. The {\it Chandra} ACIS-S
data reduction is discussed in more detail in Servillat et al. (in
preparation). Their modeling indicates that the data likely suffer
from mild pile-up, so we include a pile-up model based on
\citet{dav01} in our spectral fits, following the guidelines in the
{\it Chandra} ABC Guide to
Pileup\footnote{http://cxc.harvard.edu/ciao/download/doc/pileup\_abc.pdf}.
The grade migration parameter (denoted by $\alpha$ in the guide,
and not to be confused with the accretion disk stress parameter)
was left as a free parameter in our initial fits.  These fits
generically favored the maximum allowed value of 1.  Our
best-fit BHSPEC parameters were mildly sensitive to variations
      in this $\alpha$, but $\alpha \sim 1$ yields results which are
      consistent with the S1 and S2 datasets.  Hence we fixed $\alpha
      = 1$ for subsequent analysis.  The frame time was set to 0.8 s
to match the observational setup,
      and all other parameters were left at their XSPEC defaults.

All spectra were rebinned to require a minimum of 20 counts per
bin.

Using
XSPEC\footnote{http://heasarc.nasa.gov/docs/xanadu/xspec/index.html}
(v12.6.0q), we fit each observation independently to determine its
suitability for modeling with BHSPEC.  Specifically, we evaluate the
level of disk dominance for typical best-fit parameters.  For both of
the {\it XMM-Newton} observations, we fit all EPIC datasets (MOS1,
MOS2, and pn) simultaneously.  An additional hard X-ray component is
necessary to obtain a good fit for these datasets.  Fits with SIMPL
typically find a best-fit scattering fraction $f_{\rm sc} \gtrsim 0.5$
for the XMM1 data and $f_{\rm sc} \lesssim 0.15$ for XMM2.  These
results are consistent with those of \citet{far09} who find an
acceptable fit to the XMM1 data with an absorbed power-law, but
require an additional soft component \citep[modeled with
  DISKBB,][]{mit84} for the XMM2 data. Our fits suggest that the XMM1
spectrum is more typical of a steep power-law state, while the XMM2
spectrum is consistent with a thermally dominated state. A large
$f_{\rm sc}$ in the fit with SIMPL is problematic for our accretion
disk model, which does not account for the effects of irradiation by
neighboring corona.  Hence, we only report our results from the XMM2
analysis below.  We note that the XMM1 data generally favor a best-fit
$M$ that is higher than XMM2 for the same $i$.  This yields a cooler
disk model, consistent with the fact that more of the high energy flux
is accounted for with the SIMPL component in these data.  However, the
joint confidence contours on $a_*$ and $M$ are much larger for the
XMM1 data, so the best-fit values are still consistent with the XMM2
values at 90\% confidence.

The {\it Swift} and {\it Chandra} observations caught the source at
higher luminosities than the XMM2 observation, but the overall
signal-to-noise is still lower because of the lower effective area of
the {\it Swift} XRT and the short duration (10 ks) of the {\it Chandra}
ACIS-S observation.  Due to the lower signal-to-noise and stronger
disk dominance, a suitable fit is provided by the (absorbed) BHSPEC
model alone.  The addition of the SIMPL model only provides a marginal
improvement to the fit and the best-fit SIMPL parameters are poorly
constrained.  We also fit the combined S1 and S2 observations.  In
this case we tie all BHSPEC parameters together, except for $\ell$,
which is allowed to vary independently for S1 and S2.  Even for the
combined dataset, we find a good fit with BHSPEC alone, and poor
constraints on the SIMPL model parameters.  Therefore, we do not
include the SIMPL model in subsequent analysis of the combined S1 and
S2 datasets or in our analysis of the {\it Chandra} data.

\begin{figure}
\includegraphics[width=0.5\textwidth]{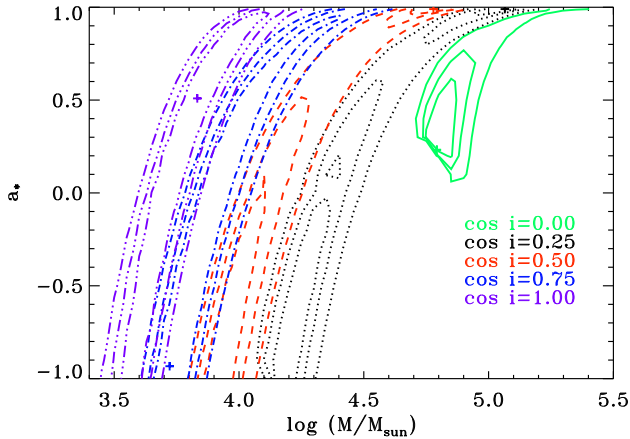}
\caption{
Joint confidence contours of $a_*$ and $M$, assuming fixed
inclination for the XMM2 data.  Each set of contours is computed
by fitting the BHSPEC model with fixed $\cos i= 0$ (solid, green),
0.25 (dotted, black), 0.5 (dashed, red), 0.75 (dot-dashed, blue), or 1
(triple-dot-dashed, violet).  The three contours in each set are
$\Delta \chi^2=2.3$, 4.61, and 9.21, corresponding to 68\%, 90\%, and
99\% joint confidence on $a_*$ and $\log M$.
\label{f:spin}}
\end{figure}

\section{Results}
\label{res}

\subsection{{\it XMM-Newton} Results}

We first consider the XMM2 observation.  The soft thermal
component is largely characterized by only two parameters: the energy
at which the spectrum peaks and the overall normalization.  In
practice, this leads to degeneracies in the best-fit parameters of
BHSPEC \citep[see e.g.][]{ddb06}, unless all but two of the parameters
can be independently constrained.  Since we can only constrain $D$
independently, we will need to consider joint variation of the
remaining parameters.

\begin{figure}
\includegraphics[width=0.5\textwidth]{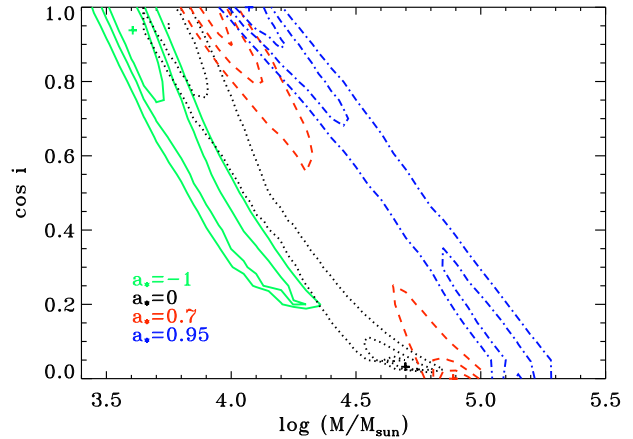}
\caption{
Joint confidence contours for BH mass and inclination, assuming fixed
BH spin, for the XMM2 observation.  Each set of contours is computed
by fitting the BHSPEC model with fixed $a_*= -1$ (solid, green), 0
(dotted, black), 0.7 (dashed, red), or 0.95 (dot-dashed, blue).  The
three contours in each set correspond to 68\%, 90\%, and 99\% joint
confidence on $\cos i$ and $\log M$.
\label{f:inc}}
\end{figure}

For completeness, our best-fit parameters for various choices of $i$
or $a_*$ are summarized in Table~\ref{t:xmm}.  We find acceptable fits
for all $i$.  The best-fit model and data for $i=0$ are plotted
in Figure~\ref{f:spec}. We report the 90\% confidence interval for a
single parameter, but due to the significant degeneracies in the
model, the joint confidence contours better illustrate the actual
parameter uncertainties.  Hence they are the focus of this work.

Since $M$ and $a_*$ are of primary interest to us, we first examine
their joint confidence contours, and consider the joint confidence of
$M$ and $i$ in Section~\ref{disc}.  For illustration, it is useful to
consider several sets of contours for different choices of (fixed)
$i$, leaving $\ell$ as a free parameter. These best-fit joint confidence
contours are shown in Figure~\ref{f:spin}.  We consider five choices
for $\cos i$, evenly spaced from 0 to 1.  Each set of contours has
three curves corresponding to 68\%, 90\%, and 99\% confidence.  These
are determined by the change in $\chi^2$ relative to the best-fit
values listed in Table~\ref{t:xmm}.

Even for fixed $i$, there is a clear correlation of $a_*$ and $M$. At
99\% confidence, the entire range of $a_*$ ($-1 < a_* < 0.99$) is
allowed, and the corresponding $M$ varies by a factor of 4-5 over this
range.  The only exception is for nearly edge on systems ($\cos i \sim
0$), for which low spins are disallowed. The correlation of best-fit
$M$ with $i$ is very strong as well, consistent with previous modeling
of other ULX sources \citep{hk08}.  
This is illustrated more clearly in Figure~\ref{f:inc}, which shows
the joint confidence contours of $M$ and $\cos i$ for fixed $a_*=-1$,
0, 0.7, and 0.95.  The corresponding best-fit values are listed in
Table~\ref{t:xmm}.  At fixed $a_*$, there is up to a factor of 10
change in $M$ as inclination varies from $0^\circ$ to $90^\circ$.  The
combined overall uncertainty in the mass is nearly a factor of 100.

Despite these degeneracies, we can still place a firm lower limit of
$M \gtrsim 3000 \Msun$ on the mass of the BH in HLX-1. The limiting
mass is obtained for the case of a maximally spinning BH with a
counter-rotating disk viewed nearly face-on.  An independent lower
limit on the mass may be obtained under the assumption that the source
luminosity does not exceed the Eddington limit.  This gives the same
answer, $M \gtrsim 3000 \Msun$, which is purely coincidental.  At the
high mass end, the limit we obtain from our fits is $M \lesssim 3
\times 10^5 \Msun$, where the limiting value is obtained for nearly
maximal spins in the case where disk rotation is spin-aligned and the
system is viewed nearly edge on.  However, edge on systems may be
ruled out by the lack of X-ray eclipses if one assumes the accreting
matter is provided by a binary companion.  If one limits $i\lesssim
75^\circ$, one obtains $M \lesssim 10^5 \Msun$.  The above limits on
the mass place HLX-1 in the IMBH regime, though the very
highest masses in our allowed range approach the lower end of the mass
distribution inferred in active galactic nuclei \citep[e.g.,][]{gh07}.

\begin{figure}
\includegraphics[width=0.5\textwidth]{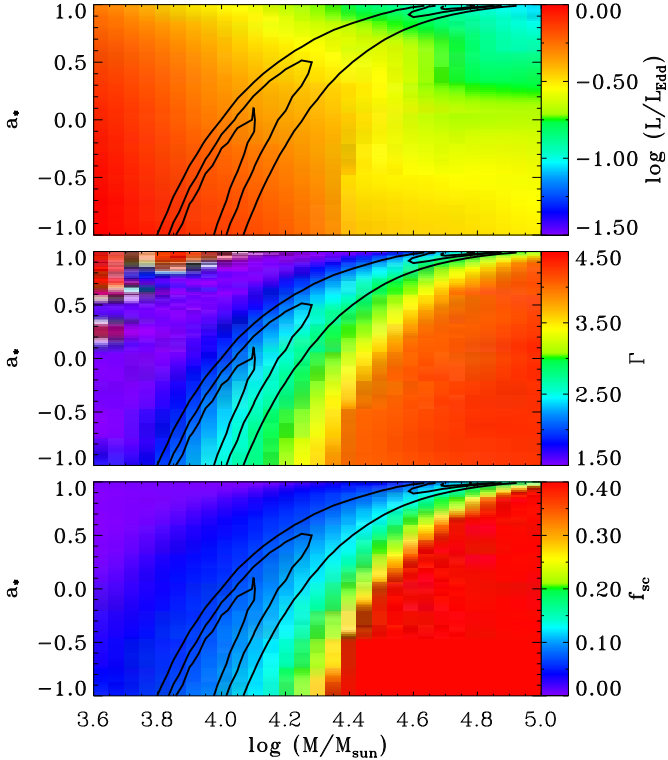}
\caption{ 
From top to bottom, the best fit Eddington ratio,
power law index, and scattering fraction in the $a_* - \log M$ plane,
assuming $\cos i=0.5$. For comparison, we overplot three contours
corresponding to 68\%, 90\%, and 99\% joint confidence on $a_*$ and
$\log M$.
\label{f:three}}
\end{figure}

We plot the variation of the other free parameters in
Figure~\ref{f:three} for the specific example of $\cos i=0.5$.  The
variation of $\ell$ is the most interesting, as there are fairly clear
correlations with $M$ and $a_*$.  Following the best-fit contour,
$\ell$ decreases as $M$ and $a_*$ increase.  This correlation arises
because $\ell$ has a role both in determining the location of the
spectral peak as well as its overall normalization, a point we
elaborate on in Section~\ref{disc}.

The bottom two panels show the variation of the SIMPL parameters.  In
contrast to $\ell$, these parameters tend to vary across (rather than
along) the confidence contours, indicating that they are highly
correlated with $\chi^2$.  A similar behavior is present in the
absorption column.  For the best-fit values we find $N_{\rm H} \simeq
1-4 \times 10^{20} \rm \; cm^{-2}$, $\Gamma \simeq 2.2-2.7$, and a
scattered fraction $\lesssim 12\%$.  The allowed $N_{\rm H}$ range is
roughly equivalent to or slightly greater than the Galactic value
\citep[$1.7 \times 10^{20} \; \rm cm^{-2}$,][]{kal05}.  The ranges for
the two SIMPL parameters are generally consistent with fits to the
thermal state of Galactic BH binaries and confirm that the bolometric
luminosity of the best-fit models is dominated by the BHSPEC
component.  For typical parameters within the confidence contours, the
accretion disk accounts for 80-95\% of the unabsorbed model
luminosity from 0.3 to 10 keV.  Comparison with Table~1 of
\citet{sd06} shows that this is a much higher disk fraction than
typically inferred in other bright ULXs, in which values 10-50\% are
more typical.

\begin{figure}
\includegraphics[width=0.5\textwidth]{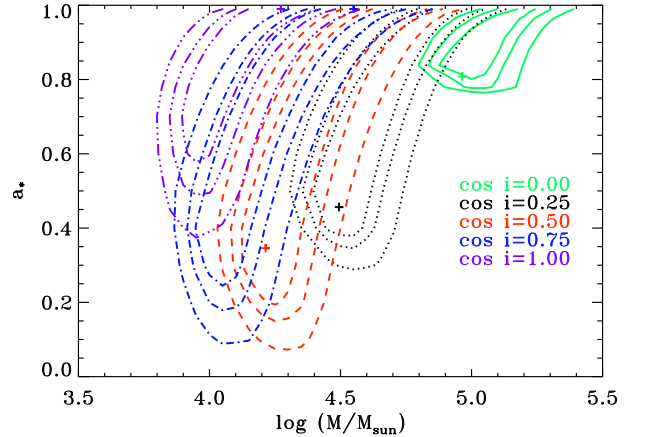}
\caption{
Joint confidence contours of $a_*$ and $M$, assuming fixed
inclination for the {\it Swift} data.  Each set of contours is
computed by fitting the BHSPEC model with fixed $\cos i= 0$ (solid,
green), 0.25 (dotted, black), 0.5 (dashed, red), 0.75 (dot-dashed,
blue), or 1 (triple-dot-dashed, violet).  The three contours in each
set are $\Delta \chi^2=2.3$, 4.61, and 9.21, corresponding to 68\%,
90\%, and 99\% joint confidence on $a_*$ and $\log M$.
\label{f:spin_swift}}
\end{figure}

\subsection{{\it Swift} and {\it Chandra} Results}

\begin{figure}[h]
\includegraphics[width=0.5\textwidth]{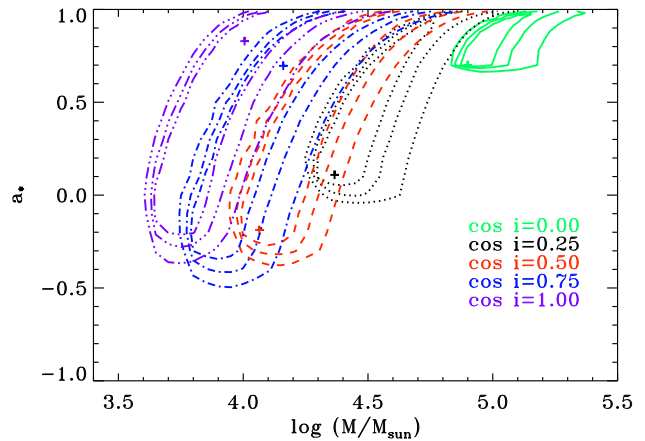}
\caption{
Joint confidence contours of $a_*$ and $M$, assuming fixed
inclination for the {\it Chandra} data.  Each set of contours is
computed by fitting the BHSPEC model with fixed $\cos i= 0$ (solid,
green), 0.25 (dotted, black), 0.5 (dashed, red), 0.75 (dot-dashed,
blue), or 1 (triple-dot-dashed, violet).  The three contours in each
set are $\Delta \chi^2=2.3$, 4.61, and 9.21, corresponding to 68\%,
90\%, and 99\% joint confidence on $a_*$ and $\log M$.
\label{f:spin_chandra}}
\end{figure}

In Figure~\ref{f:spin_swift}, we plot the best-fit joint confidence
contours of $a_*$ and $M$ for the combined {\it Swift} (S1 and S2)
data sets.  Figure~\ref{f:spin_chandra} is the corresponding plot for
the {\it Chandra} dataset.  As discussed in Section~\ref{data}, we do
not include a SIMPL component in fitting these datasets and we tie all
the BHSPEC parameters together, except $\ell$, which is allowed to
vary independently for each observation.  The corresponding best-fit
parameters are depicted with crosses and summarized in
Tables~\ref{t:swift} and \ref{t:chandra}, where $\ell_1$ and
  $\ell_2$ correspond to the S1 and S2 observations, respectively.
The reduced $\chi^2$ for the {\it Chandra} data indicate a slightly
poorer fit than with the {\it XMM-Newton} or {\it Swift} data.  This
is primarily due to narrow residuals near 0.6, 1.1 and 1.3 keV in fits
to the ACIS-S data.  Since these residuals are not highly significant
and their presence does not significantly impact the best-fit BHSPEC
parameters, we do not attempt to model them with additional
components.  Further discussion can be found in Servillat et al. (in
preparation).

\begin{deluxetable*}{lcccccc}
\tablecolumns{7}
\tablecaption{Best-Fit {\it Swift} Parameters\label{t:swift}}
\tablewidth{0pt}
\tablehead{
\colhead{$\cos i$} &
\colhead{$a_*$} &
\colhead{$\log M$\tablenotemark{a}} &
\colhead{$\log \ell_1$} & 
\colhead{$\log \ell_2$} &                                      
\colhead{$N_H$\tablenotemark{b}} &
\colhead{$\chi^2$/d.o.f.}
}
\startdata
$0$ &
$0.81^{+0.18}_{-0.02}$ &
$4.96^{+0.30}_{-0.08}$ &
$-0.00^{+0}_{-0.68}$ &
$-0.05^{+0.03}_{-0.58}$ &
$3.2^{+3.1}_{-2.6}$ &
$58.3/75$ \\
$0.25$ &
$0.46^{+0.46}_{-0.07}$ &
$4.49^{+0.38}_{-0.08}$ &
$-0.01^{+0.01}_{-0.92}$ &
$-0.06^{+0.03}_{-0.56}$ &
$5.1^{+3.1}_{-2.6}$ &
$59.6/75$ \\
$0.5$ &
$0.35^{+0.64}_{-0.22}$ &
$4.21^{+0.63}_{-0.08}$ &
$-0.02^{+0.02}_{-0.54}$ &
$-0.06^{+0.04}_{-0.42}$ &
$5.3^{+3.5}_{-2.7}$ &
$59.8/75$ \\
$0.75$ &
$0.99^{+0}_{-0.25}$ &
$4.54^{+0.09}_{-0.46}$ &
$-0.29^{+0.29}_{-0.04}$ &
$-0.33^{+0.31}_{-0.03}$ &
$5.3^{+3.5}_{-2.7}$ &
$58.9/75$ \\
$1$ &
$0.99^{+0}_{-0.05}$ &
$4.27^{+0.05}_{-0.58}$ &
$-0.03^{+0.03}_{-0.06}$ &
$-0.08^{+0.08}_{-0.06}$ &
$7.0^{+3.2}_{-3.6}$ &
$58.5/75$ \\

\enddata
\tablenotetext{a}{$M$ is in units of $\Msun$.}
\tablenotetext{b}{$N_{\rm H}$ is in units of $10^{20} \; cm^{-2}$.}
\tablecomments{
Parameters tabulated without errors were fixed during the fit.  Errors
are computed using $\Delta \chi^2=2.706$ (90\% confidence for one
parameter).  Joint confidence contours are shown in
Figure~\ref{f:spin_swift}. In these joint fits to the S1 and S2 data
sets all BHSPEC parameters are fit simultaneously to both spectra,
except $\ell$ which is fit independently.  Hence, $\ell_1$ and $\ell_2$
correspond to the S1 and S2 observations, respectively.}
\end{deluxetable*}

The {\it Swift} and {\it Chandra} data seem to place much tighter
overall constraints on $a_*$ and $M$, with a minimum $a_*$ at each
$i$.  The overall minima of $M \sim 6000 \Msun$ for {\it Swift}) and
and $M \sim 4000 \Msun$ for {\it Chandra} both correspond to
$i=0^\circ$.  Fits with retrograde accretion flows ($a_* < 0$) are
inconsistent with the {\it Swift} data and models with $a_* \lesssim
-0.5$ are ruled out by the {\it Chandra} spectrum.  For values of
$a_*$ allowed by both data sets, the position of the confidence
contours for a given $i$ are consistent with each other.  They are
also broadly consistent with the XMM2 results, although offset to
slightly higher $M$.  However, due to the lower signal-to-noise of the
{\it Swift} and {\it Chandra} datasets, the range of allowed $M$ for a
given $i$ is larger than in Figure \ref{f:spin}.

The differences between the {\it XMM-Newton} fits and the {\it Swift}
or {\it Chandra} results at low (negative) $a_*$ are driven primarily
by the larger luminosities of the disk component in the {\it
  Swift} and {\it Chandra} observations.  With the exception of nearly
edge on systems ($\cos i =0$) we find $\ell < 1$ within the 99\%
confidence contours in Figure~\ref{f:spin}.  In the {\it Swift} and
{\it Chandra} data, $\ell \ge 1$ is required for all $i$, at
  sufficiently low $a_*$.  As we discuss further in
Section~\ref{disc}, reductions in $a_*$ and $M$ must be offset by
increases in $\ell$, but our models are capped at $\ell=1$, which sets
a lower limit on the allowed $a_*$ and $M$ for a given choice of $i$.
Hence, in contrast to the XMM2 observation, these tighter constraints
depend strongly on the assumption that luminosity does not exceed the
Eddington limit: $\ell \le 1$.

We also fit the combined {\it Swift}, {\it Chandra} and {\it
  XMM-Newton} datasets assuming no correction was needed to the
relative effective area of the various instruments. In principle, a
constant offset could be included based on cross-calibration analysis
\citep[e.g.,][]{tsu11}, but even without such a correction, we
obtain good fits to the combined dataset.  The range of allowed
$a_*$ is similar to that found with the {\it Swift} data alone, but
the width of the contours are comparable to those from the XMM2 fits
plotted in Figure~\ref{f:spin}.  At 99\% confidence, the minimum
allowed $M$ was about $6000 \Msun$, again for $i=0^\circ$ and $\ell =
1$.

\section{Discussion}
\label{disc}
 
Due to the complex interplay between the frequency and angular
distribution of emitted photons in the rest frame of the flow, and the
relativistic effects on the disk structure and radiation, it is
preferable to use a self-consistent relativistic model to draw
quantitative conclusions.  Nevertheless, one can understand the basic
trends and correlations with a simpler model, in which the relativistic
effects are encapsulated into two parameters, similar in spirit to the
pioneering work of \citet{zcc97}. 

\subsection{A Minimal Relativistic Disk Model}

To first approximation, thermal state disk spectra are characterized
by only two parameters: a peak energy and an overall normalization.
Indeed the popular DISKBB model \citep{mit84}, with only two
  parameters, provides an acceptable fit to the soft thermal component
  of the XMM2 data \citep{far09}.  Although we obtain all our
constraints by fitting a relativistic accretion disk model
directly to the spectrum, our fit results are easily understood in
terms of these two constraints.  These observables can be mapped onto
a characteristic color temperature $T_{\rm obs}$ ($\approx T_{\rm
    in}$ from DISKBB), and since the distance $D$ is known, a
characteristic luminosity $L_{\rm obs}$.

\begin{deluxetable}{lccccc}
\tablecolumns{6}
\tablecaption{Best-Fit {\it Chandra} Parameters\label{t:chandra}}
\tablewidth{0pt}
\tablehead{
\colhead{$\cos i$} &
\colhead{$a_*$} &
\colhead{$\log M$\tablenotemark{a}} &
\colhead{$\log \ell$ } & 
\colhead{$N_H$\tablenotemark{b}} &
\colhead{$\chi^2$/d.o.f.}
}
\startdata
$0$ &
$0.70^{+0.27}_{-0.05}$ &
$4.90^{+0.28}_{-0.04}$ &
$-0.01^{+0.01}_{-0.63}$ &
$0.0^{+2.0}_{-0.0}$ &
$48.4/38$ \\
$0.25$ &
$0.11^{+0.85}_{-0.07}$ &
$4.37^{+0.37}_{-0.06}$ &
$-0.01^{+0.01}_{-0.73}$ &
$1.3^{+3.5}_{-1.3}$ &
$50.4/38$ \\
$0.5$ &
$-0.19^{+1.18}_{-0.10}$ &
$4.06^{+0.77}_{-0.06}$ &
$0.00^{+0.00}_{-0.68}$ &
$1.8^{+4.0}_{-1.8}$ &
$50.1/38$ \\
$0.75$ &
$0.70^{+0.29}_{-1.06}$ &
$4.16^{+0.54}_{-0.39}$ &
$-0.25^{+0.25}_{-0.20}$ &
$2.3^{+4.8}_{-2.3}$ &
$49.4/38$ \\
$1$ &
$0.83^{+0.16}_{-0.16}$ &
$4.01^{+0.32}_{-0.26}$ &
$-0.16^{+0.16}_{-0.20}$ &
$2.0^{+5.8}_{-2.0}$ &
$49.9/38$ \\

\enddata
\tablenotetext{a}{$M$ is in units of $\Msun$.}
\tablenotetext{b}{$N_{\rm H}$ is in units of $10^{20} \; cm^{-2}$.}
\tablecomments{
Parameters tabulated without errors were fixed during the fit.  Errors
are computed using $\Delta \chi^2=2.706$ (90\% confidence for one
parameter).  Joint confidence contours are shown in
Figure~\ref{f:spin_chandra}.}
\end{deluxetable}

The characteristic temperature is related to the peak effective
temperature $\teff$ of the accretion disk.  For illustration, we
assume this is equivalent to the effective temperature at the inner
edge of the disk.  Ignoring relativistic terms and constants of order
unity, the flux near the inner edge of an accretion disk can be
approximated by
\begin{equation}
\sigma \teff^4 \simeq \frac{G M \Mdot}{R_{\rm in}^3,}
\label{eq:diskflux}
\end{equation}
where $G$ is Newton's constant, and $\Mdot$ is the accretion rate.
The inner radius of the disk $R_{\rm in}$ corresponds to the innermost
stable circular orbit (ISCO) in our model.

Due to deviations from blackbody emission and relativistic effects on
photon propagation, $T_{\rm obs} \neq \teff$.  We parametrize the
deviations from blackbody via a spectral hardening factor (or color
correction) $f_{\rm col}$ and the relativistic energy shifts with
$\delta$ so that $T_{\rm obs} = f_{\rm col} \delta \teff$.  In
practice, $\delta$ is primarily a function of $i$ and $a_*$ with
$\delta \gtrsim 1$ typical for most $a_*$ and $i$ in our models
(i.e. Doppler blueshifts are generally more important than Doppler
and general relativistic redshifts).  In contrast, $f_{\rm col}$ is
dependent on $M$, $a_*$, and $\ell$ and largely independent of
$i$. Using this parametrization and equation (\ref{eq:diskflux}) we
obtain
\begin{equation}
T_{\rm obs} \simeq T_0 f_{\rm col} \delta \left(\frac{\ell}{r_{\rm in}^2 m}\right)^{1/4},
\label{eq:tobs}
\end{equation}
where, $m =M/\Msun$, $r_{\rm in} =R_{\rm in}/R_g$, $T_0 =
(c^5 / G \Msun \kappa_{\rm es} \sigma)^{1/4}$, $R_g=GM/c^2$ is the
gravitational radius, and $\kappa_{\rm es}$ is the electron scattering
opacity.  Here, $\ell =\eta \Mdot c^2/L_{\rm Edd}$, and for simplicity
we have approximated $\eta \sim 1/r_{\rm in}$.

Using the above definitions the observed luminosity is
\begin{equation}
L_{\rm obs} \simeq L_0 l m \mu,
\label{eq:lobs}
\end{equation}
where $L_0 \equiv 4\pi G \Msun c/\kappa_{\rm es}$ is the Eddington
luminosity for a one solar mass BH and $\mu$ is a variable that
encapsulates all of the angular dependence of the radiation field.  In
an isotropically emitting, Newtonian disk $\mu = \cos i$, accounting
for the inclination dependence of the disk projected area.  In our
models, two other effects are important as well: electron scattering
and relativistic beaming which tend to make the disk emission more
limb-darkened and limb-brightened, respectively.  The latter depends
significantly on the spin so $\mu$ can be a strong function of $a_*$
as well as $i$.  Lensing by the BH can also be important, particularly
for nearly edge on systems where it places a lower limit on the
effective projected area of the disk.

\begin{figure}
\includegraphics[width=0.5\textwidth]{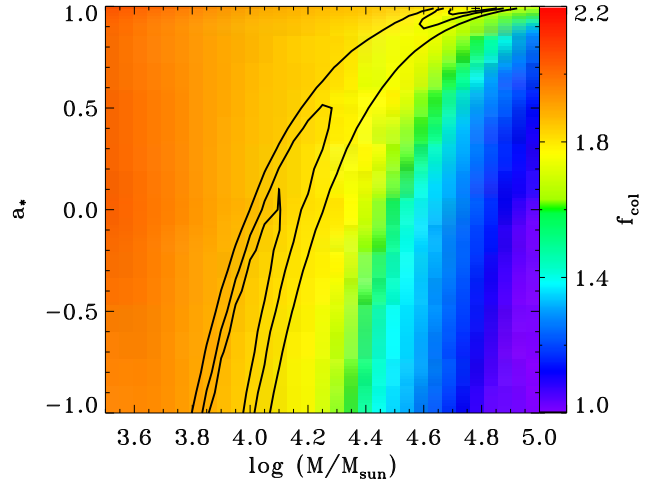}
\caption{ 
Color correction corresponding to the best-fit model in the $a_* -
\log M$ plane, assuming $\cos i=0.5$. For comparison, we overplot
three contours corresponding to 68\%, 90\%, and 99\% joint confidence
on $a_*$ and $\log M$.
\label{f:fcol}}
\end{figure}

Equations (\ref{eq:tobs}) and (\ref{eq:lobs}) can be solved for $m$
and $\ell$, yielding
\begin{equation}
M \simeq \Msun \left(\frac{T_{\rm obs}^4 L_0}{T_0^4 L_{\rm obs}} \right)^{1/2}
\frac{f_{\rm col}^2 \delta^2}{\mu^{1/2} r_{\rm in}}.
\label{eq:mass}
\end{equation}
The quantities in parentheses on the right hand side of equation
(\ref{eq:mass}) are observational constraints or constants of nature.
Uncertainties in $D$ enter through $L_{\rm obs} \propto D^2$, so we
approximately have $M \propto D^{-1}$.  Therefore, the relatively
small uncertainty in $D$ contributes only a very modest additional
uncertainty in $M$.  The remaining quantities are functions of model
parameters: $r_{\rm in}(a_*)$, $\mu(a_*, i)$, $\delta(a_*,i)$, and
$f_{\rm col}(a_*,\ell,M)$.

\subsection{Correlations among the Best-fit Parameters}

The dependence of $r_{\rm in}$ on $a_*$ is the primary driver of the
correlation of $M$ with $a_*$ in Figure~\ref{f:spin}.  As $a_*$
increases from -1 to 0.99, $r_{\rm in}$ decreases from 9 to 1.45,
driving corresponding increases in $M$.  This variation is strongest
as $a_* \rightarrow 1$, leading to the ``bend'' in the confidence
contours at high $a_*$.  Both $\delta$ and $f_{\rm col}$ increase with
$a_*$, and contribute to the correlation as well.

The spectrum in BHSPEC is calculated directly, so there is no explicit
$f_{\rm col}$, but we can estimate $f_{\rm col}$ by taking the best
fit BHSPEC models and fitting them with the KERRBB model, for which
$f_{\rm col}$ is a parameter.\footnote{Specifically, we fix $a_*$,
  $i$, $M$ to correspond to the BHSPEC parameters, leaving $f_{\rm
    col}$ and $\Mdot$ as free parameters in the KERRBB fit.}  The
variation of $f_{\rm col}$ is shown for $\cos i=0.5$ in Figure
\ref{f:fcol}.  Although $f_{\rm col}$ can vary by more than a factor 2
over the parameter range of interest, its change within the best fit
contours is significantly less.  For $\cos i \sim 0.5$, $\delta$
increases modestly with $a_*$ due to Doppler blueshifting, so the
product of $f_{\rm col}$ and $\teff$ must decrease to maintain
agreement with $T_{\rm obs}$. Since $\teff$ is the primary driver of
variations in $f_{\rm col}$, and $f_{col}$ and $\teff$ are positively
correlated, both $\teff$ and $f_{\rm col}$ must decrease as $a_*$
increases.  The same argument applies outside the best-fit contours,
but leads to a much larger variation in $f_{\rm col}$.

The confidence contours in Figure~\ref{f:inc} confirm a strong
anti-correlation of $M$ with $\cos i$, or equivalently, a correlation
of $M$ with $i$.  This correlation is driven primarily through the $i$
dependence of $\delta$ and $\mu$.  As $i$ increases, the projected
area of the disk decreases while limb darkening shifts intensity to
lower $i$. The combined effects lead to a significant decrease in
$\mu$ as $i$ increases. For higher $a_*$, these effects are somewhat
mitigated by relativistic beaming, which tends to shift intensity to
larger $i$.  In addition, Doppler blue-shifts due to the Keplerian
motion cause $\delta$ to increase with $i$.  Both effects are present
and drive a positive correlation of $M$ and $i$, consistent with
Figure \ref{f:inc}.  The anti-correlation of $i$ and $\mu$ dominates
at low spin, while the correlation of $\delta$ with $i$ dominates at
higher spin because rotational velocities are a large fraction of the
speed of light.  The effects contribute comparably for $a_* \sim 0.7$.

We note that \citet{hk08} found a similar correlation in their
analysis of several ULX sources.  To explain their results, they
attribute the correlation to Doppler shifts (i.e. the dependence of
$\delta$ on $i$), which is consistent with the fact that their
best-fit models favored high $a_*$.

The anti-correlation of best-fit $\ell$ with $M$ and $a_*$ follows
directly from equations (\ref{eq:tobs}) and (\ref{eq:lobs}).
As $M$ increases, $\ell$ must decrease to keep $L_{\rm obs}$
approximately constant, but keeping $T_{\rm obs}$ constant then
requires an increase in $a_*$ (i.e. a reduction in $r_{\rm in}$).
This is why high $M$ and low $a_*$ (and vice-versa) yield poor fits
for $\cos i=0.5$ in Figure \ref{f:three}.  

With this understanding of how $M$, $i$, $\ell$, and $a_*$ correlate,
it is useful to consider what ultimately sets the minimum allowed $M$
in the two data sets.  The $L_{\rm obs}$ constraint
(eq. [\ref{eq:lobs}]) allows $M$ to decrease as long as there is a
corresponding increase in $\mu$ or an increase in $\ell$.  For fits to
the XMM2 data, $\ell$ is set by the $T_{\rm obs}$ constraint
(eq. [\ref{eq:tobs}]).  Since $\ell$ increases as $M$ decreases,
$r_{\rm in}$ must increase.  Hence, the maximum $\ell$ and minimum $M$
are obtained for $a_*=-1$.  This argument leads to a minimum $M$ for
any fixed $i$ and since $\mu$ is maximum for face on disks, the
`global' minimum of $M$ occurs at $i=0^\circ$ and corresponds to
$\ell$ slightly less then unity.  Hence, the constraints that $M
\gtrsim 3000 \Msun$ is {\it independent} of the argument that
$\ell < 1$.

For the {\it Swift} and {\it Chandra} datasets, which consist of
observations with higher $L_{\rm obs}$, this is not the case.  Our
models are capped at $\ell=1$ due to inconsistencies in the underlying
model assumptions for $\ell \gtrsim 1$.  For some minimum $M$ ($M \sim
6000 \Msun$: {\it Swift}; $M \sim 4000 \Msun$: {\it Chandra}), equation
(\ref{eq:lobs}) cannot be satisfied for $\ell \le 1$, even with
$i=0^\circ$.  This corresponds also to a lower limit on $a_*$ since
further increases in $r_{\rm in}$ cannot be offset by decreases
(increases) in $M$ ($\ell$) to keep $T_{\rm obs}$ constant.  If we
ignored the internal inconsistencies in thin disk assumptions and
extended our models to $\ell > 1$, we expect we would obtain
reasonable fits for lower $M$ and $a_*$ as equations (\ref{eq:tobs})
and (\ref{eq:lobs}) could still be satisfied.  Hence, the more
stringent lower limit on $M$ and $a_*$ for the {\it Swift} and {\it
  Chandra} fits is not independent of the Eddington limit.

For the XMM2 data, the minimum allowed $M$ is reached for models with
$\ell \sim 0.7$, just below the Eddington limit.\footnote{Our $\ell$
  only accounts for the luminosity of the BHSPEC component, the total
  bolometric luminosity $L_{\rm bol}$ of the systems is larger once the coronal
  component is added.  For typical SIMPL parameters, the coronal
  component accounts for $\lesssim 25$\% of the total model flux, even
  after extrapolation to larger energies.  Hence, $L_{\rm bol}/L_{\rm
    Edd} \lesssim 1$ still holds.} The assumptions underlying the thin
accretion disk model are likely to break down for $\ell \sim 1$
although it is difficult to precisely estimate the $\ell$ value where
our spectral constraints are no longer reasonable.  General
relativistic magnetohydrodynamic simulations
\citep[e.g.,][]{pen10,kul11} and slim disk models \citep{sad10}
suggest that the thin disk spectral models remain reliable to (at
least) $\ell \lesssim 0.3$.  Observations of Galactic X-ray binaries
suggest that there are no abrupt changes at this $\ell$ \citep{ste10},
so it is plausible that models remain basically sound even for
somewhat higher $\ell$.

The strong correlation of $\chi^2$ with $\Gamma$ and $f_{\rm sc}$ in
Figure~\ref{f:three} (as well as a similar one for $N_{\rm H}$)
results from the failure of the BHSPEC model to adequately approximate
the thermal emission for $M$ and $a_*$ outside the confidence
contours.  If BHSPEC is a poor match to the thermal spectrum, the fit
compensates by adjusting $N_{\rm H}$ or the SIMPL parameters.  For
example, at low $M$ and $a_*$, the disk is too cold to match the XMM2
data, so SIMPL adjusts by increasing the scattering fraction and
making the power law steeper, to better fit the high energy tail of
the thermal emission.  To prevent an excess at lower energies, the
$N_{\rm H}$ increases simultaneously.  For the lower signal-to-noise
{\it Swift} and {\it Chandra} data, $N_{\rm H}$ adjusts in a similar
manner, even though the SIMPL component is absent. The width of the
confidence contours appears to be set largely by the effectiveness of
these compensation mechanisms.  Hence, the extent of the confidence
contours for a given $i$ could presumably be reduced with better
statistics at high energies to constrain the SIMPL parameters and
independent constraints on $N_{\rm H}$.

Finally, it is worth considering the degree to which our correlations
could be reproduced by a non-relativistic analysis, but assuming
$r_{\rm in}$ is equal to the ISCO radius.  For example, assuming
constant $f_{\rm col} \sim 1.7$, $\delta \sim 1$, and $\mu \sim \cos
i$ in equation \ref{eq:mass}, one can recover some aspects of the
inferred correlations of $M$ with $a_*$ and $i$.  The sensitivity of
$M$ to $a_*$ would be well approximated for low-to-moderate $a_*$, but
the dependence of $\delta$, $f_{\rm col}$, and $\mu$ on $a_*$ can lead
to modest discrepancies as $a_* \rightarrow 1$. In contrast, the
correlation of $M$ with $i$ would only be crudely reproduced since the
above prescription would suggest $M \propto (\cos i)^{-1/2}$.  This
underestimates the sensitivity of $M$ to $i$ for low to moderate $i$,
but overestimates it as $i \rightarrow 90^\circ$, where the projected
area goes to zero in a non-relativistic model.  Since the low to
moderate range is probably more relevant for observed systems, the
overall uncertainty of $M$ would be underestimated.  We also note that
although the assumption of a constant $f_{\rm col} \sim 1.7$ turns out
to be approximately correct ($f_{\rm col} \approx 1.8 \pm 0.1$ in
Figure~\ref{f:fcol}), it was not clearly justified a priori and may
not be as good of an assumption for other ULX sources.  In principle,
one could improve this analysis by estimating $f_{\rm col}$, $\mu$,
and $\delta$ from BHSPEC (or some similar model), but at that level of
sophistication, it seems more sensible to fit the relativistic model
directly.

\subsection{Sensitivity to BHSPEC Model Assumptions}

There are a number of assumptions present in the BHSPEC model that
could have some impact on $f_{\rm col}$. In particular, magnetic
fields (and associated turbulence) may play a role in modifying the
disk vertical structure and radiative transfer
\citep[e.g. see][]{dav05, dh06,bla06,dav09}.  Another assumption of
interest is our choice of $\alpha=0.01$.  For $\alpha \lesssim 0.01$,
the models depend very weakly on $\alpha$ for the parameter range
relevant to our fit results.  For higher values of $\alpha$, the
typical color correction is larger.  For low to moderate $\ell$ and
$a_*$, $f_{\rm col}$ increases by less than 25 \% as $\alpha$
increases from 0.01 to 0.1. Much larger shifts can occur if both
$\ell$ and $a_*$ are larger ($a_* \gtrsim 0.8$ and $\ell \gtrsim 0.3$;
\citealt{dd08}), but models in this range overpredict $T_{\rm obs}$
and are irrelevant to our results.  Hence, if the characteristic
$\alpha$ associated with real accretion flows is larger (as some
models of dwarf novae and some numerical simulations suggest,
\citealt{kpl07}), the effect would be to shift our best-fit contours
to higher $M$, but only by a modest amount.  For the parameters
corresponding to the lower $M$ limit ($i=0^\circ$, $\ell = 0.7$, and
$a_* = -1$), BHSPEC yields $f_{\rm col} \sim 2$.  From equation
(\ref{eq:mass}) we see that reducing $f_{\rm col}=1$ (the absolute
minimum) only reduces $M$ by a factor of 4, still placing HLX-1 in the
IMBH regime.

Alternatively, one could make the disk around a low $M$ BH look cooler
by truncating it at larger radius.  Equation (\ref{eq:mass}) suggests
that decreasing $M$ to a value near $30 \Msun$ would require a factor
of 100 increase to $r_{\rm in} \sim 900$.  Such an interpretation
would need to explain why the flow does not radiate inside this
radius.  Since the energy does not come out in the hard X-rays, it
cannot be a transition to an advection dominated accretion flow, which
is often invoked to explain the low state of Galactic X-ray binaries
\citep[e.g.,][]{esi01}.  Furthermore, since $\eta \sim 1/r_{\rm in}$,
the required $\Mdot$ would increase by a factor of 100 to $\Mdot \sim
10^{-2} \; \Msun \; \rm yr^{-1}$.  For $\Mdot \sim 10^{-4} \; \Msun \;
\rm yr^{-1}$, the accretion rate and time variability of HLX-1 present
a challenge to standard models of mass transfer \citep{las11}.  Hence,
it is unlikely that such a high rate is even feasible in a binary mass
transfer scenario.

Finally, one could plausibly obey the Eddington limit by assuming a
large beaming factor \citep[e.g.][]{kin01,fre06,kin08} so that $L_{\rm
  obs} \gg L_{\rm iso}$, which (in this formalism) is equivalent to
increasing $\mu$ for some narrow range of $i$.  Obeying the Eddington
limit with $M \sim 30 \Msun$ would require $\mu \sim 100$, but note
that this is insufficient for explaining $T_{\rm obs}$ due to the
$\mu^{1/2}$ dependence in equation (\ref{eq:mass}).  Fixing $\ell =1$
and decreasing $M$ by a factor of 100 yields a factor of 3
increase in $T_{\rm obs}$.  Maintaining agreement with $T_{\rm obs}$
requires $M$ and $\ell$ to decrease proportionately, which requires a
factor of 1000 increase in $\mu$.  Any scenario with such large
beaming factors probably requires a relativistic outflow or very
different accretion flow geometry so these simple scalings may not
strictly apply.  Nevertheless, we emphasize that explaining the soft
emission in HLX-1 presents a serious challenge to any beaming
model, but is naturally explained by the IMBH interpretation.

\section{Summary and Conclusions}
\label{conc}

Using BHSPEC, a fully-relativistic accretion disk model, we fit
several disk dominated observations of HLX-1 for which the luminosity
exceeds $10^{42}$~$\rm erg \; s^{-1}$.  Due to degeneracies in the
best-fit model parameters, 90\% joint confidence uncertainties are
rather large, yielding a factor of 100 uncertainty in the best-fit BH
mass.  For fits to the {\it XMM-Newton} data, we obtain a lower limit
of $M \gtrsim 3000 \Msun$, where the limit corresponds to $i=0^\circ$,
$a_*=-1$, and $\ell =0.7$.  We emphasize that this limit is driven by
the need to reproduce the shape and peak energy of the thermal
component in the spectrum.  Hence, the Eddington limit plays no role
in this constraint.  Constraints from fits to {\it Swift} and {\it
  Chandra} observations, which correspond to higher luminosities,
nominally offer a more restrictive lower bound of $M \gtrsim 6000
\Msun$, but this bound is subject to the Eddington limit because our
model grid is limited to a maximum luminosity $\ell=1$.

We also find an absolute upper bound of $M \lesssim 3 \times 10^5
\Msun$ with both datasets, this limit corresponding to nearly edge-on
($i=90^\circ$) disks with near maximal spins ($a_* \sim 0.99$).  This
upper limit is subject to the uncertainties in the models at very high
spin and high inclination (most notably our neglect of returning
radiation and assumption of a razor thin geometry).  The lack of X-ray
eclipses and the absence of evidence for nearly edge-on X-ray binary
systems in the Milky Way \citep{nm05} motivate a limit on $i \lesssim
75^\circ$ and, therefore, $M \lesssim 10^5$.  An argument against $i
\sim 90^\circ$ based on absence of eclipses assumes that the accreting
matter is being provided by a binary companion, but obscuration by a
flared outer disk may generically limit the range of observable $i$.
For $M \gtrsim 10^5 \Msun$, HLX-1 would be consistent with the lower
end of the mass distribution inferred in active galactic nuclei
\citep[e.g.,][]{gh07}, but would still be distinctive because of its
off-nuclear location in ESO~243-49.

Other parameters of interest, such as $i$ and $a_*$, are essentially
unconstrained by the data, unless we require that the disk must
radiate below the Eddington luminosity, in which case $a_* > 0$ or
$a_* > -0.5$ are required by the {\it Swift} and {\it Chandra} data,
respectively.  Observations with improved signal-to-noise are unlikely
to significantly tighten these $M$ constraints, as the allowed $M$
range is set primarily by uncertainties in $i$ and $a_*$.  Independent
estimates for $a_*$ and $i$ are ultimately needed to improve our $M$
constraints, and could plausibly be provided by modeling of broad Fe
K$\alpha$ lines or X-ray polarization \citep{lnm09} if such data
became available. If a broad Fe line is present in HLX-1, obtaining
the signal-to-noise necessary to resolve it would require unfeasibly
long exposure times with {\it XMM-Newton} and other existing X-ray
missions.  However, such constraints may be possible for future
missions with larger collecting areas.

In summary, despite the rather large range of $M$ allowed by the
spectral fits of HLX-1 presented here, our study strongly suggests
that the BH in HLX-1 is a genuine IMBH.

\acknowledgements{We thank the anonymous referee for useful comments
  that improved the paper.  SWD is supported in part through NSERC of
  Canada. SAF gratefully acknowledges funding from the UK Science and
  Technology Research Council and the Australian Research Council.  RN
  acknowledges support from NASA grant NNX11AE16G. MS acknowledges
  supports from NASA/Chandra grant DD0-11050X and NSF grant
  AST-0909073.  This research has made use of data obtained from the
  Chandra Data Archive and software provided by the Chandra X-ray
  Center.  Based on observations from {\it XMM-Newton}, an ESA science
  mission with instruments and contributions directly funded by ESA
  Member States and NASA.}

\end{document}